\documentclass[12pt]{article}
\usepackage{amsmath}
\usepackage{amssymb}

\allowdisplaybreaks
\begin{document}

\author{C. Bizdadea\thanks{%
e-mail address: bizdadea@central.ucv.ro}, E. M. Cioroianu\thanks{%
e-mail address: manache@central.ucv.ro}, S. C. Sararu\thanks{%
e-mail address: scsararu@central.ucv.ro} \\
Faculty of Physics, University of Craiova\\
13 A. I. Cuza Str., Craiova 200585, Romania}
\title{Couplings between a collection of BF models\\
and a set of three-form gauge fields}
\date{}
\maketitle

\begin{abstract}
Consistent interactions that can be added to a free, Abelian gauge theory
comprising a collection of BF models and a set of three-form gauge fields
are constructed from the deformation of the solution to the master equation
based on specific cohomological techniques. Under the hypotheses of smooth,
local, PT invariant, Lorentz covariant, and Poincar\'{e} invariant
interactions, supplemented with the requirement on the preservation of the
number of derivatives on each field with respect to the free theory, we
obtain that the deformation procedure modifies the Lagrangian action, the
gauge transformations as well as the accompanying algebra.

\end{abstract}

Topological field theories~\cite{birmingham91}--\cite{labastida97} are
important in view of the fact that certain interacting, non-Abelian versions
are related to a Poisson structure algebra~\cite{stroblspec} present in
various versions of Poisson sigma models~\cite{psmikeda94}--\cite%
{psmcattaneo2001}, which are known to be useful at the study of
two-dimensional gravity~\cite{grav2teit83}--\cite{grav2grumvassil02}
(for a detailed approach, see~\cite{grav2strobl00}). It is well
known that pure three-dimensional gravity is just a BF theory.
Moreover, in higher dimensions general relativity and supergravity
in Ashtekar formalism may also be
formulated as topological BF theories with some extra constraints \cite%
{ezawa}--\cite{ling}. In view of these results, it is important to know the
self-interactions in BF theories as well as the couplings between BF models
and other theories. This problem has been considered in literature in
relation with self-interactions in various classes of BF models~\cite%
{defBFizawa2000}--\cite{defBFijmpajuvi06} and couplings to matter fields~%
\cite{defBFepjc} and vector fields~\cite{defBFikeda03}--\cite%
{defBFijmpajuvi04} by using the powerful BRST cohomological reformulation of
the problem of constructing consistent interactions within the Lagrangian~%
\cite{deflag} or the Hamiltonian~\cite{defham} setting. Other aspects
concerning interacting, topological BF models can be found in~\cite%
{otherBFikeda02}--\cite{otherBFikeda06}. On the other hand, models with $p$%
-form gauge fields play an important role in string and superstring theory
as well as in supergravity. In particular, three-form gauge fields are
important due to their presence in eleven-dimensional supergravity. Based on
these considerations, the study of interactions between BF models and
three-forms appears as a topic that might enlighten certain aspects in both
gravity and supergravity theories.

The scope of this paper is to investigate the consistent interactions that
can be added to a free, Abelian gauge theory consisting of a collection of
BF models and a set of three-form gauge fields. This matter is addressed by
means of the deformation of the solution to the master equation from the
BRST-antifield formalism~\cite{deflag}. Under the hypotheses of smooth,
local, PT invariant, Lorentz covariant, and Poincar\'{e} invariant
interactions, supplemented with the requirement on the preservation of the
number of derivatives on each field with respect to the free theory, we
obtain the most general form of the theory that describes the
cross-couplings between a collection of BF models and a set of three-form
gauge fields. The resulting interacting model is accurately formulated in
terms of a gauge theory with gauge transformations that close according to
an open algebra (the commutators among the deformed gauge transformations
only close on the stationary surface of deformed field equations), which are
on-shell, second-order reducible.

Our starting point is a four-dimensional, free theory, describing a
collection of topological BF models (each of them involving two types of
one-forms, a set of scalar fields, and a system of two-forms) and a set of
Abelian $3$-form gauge fields, with the Lagrangian action
\begin{equation}
S_{0}\left[ A_{\mu }^{a},H_{\mu }^{a},\varphi _{a},B_{a}^{\mu \nu
},V_{\mu \nu \rho }^{A}\right] =\int d^{4}x\left( H_{\mu
}^{a}\partial ^{\mu }\varphi _{a}+\tfrac{1}{2}B_{a}^{\mu \nu
}\partial _{[\mu }^{\left. {}\right. }A_{\nu ]}^{a}-\tfrac{1}{2\cdot
4!}F_{\mu \nu \rho \lambda }^{A}F_{A}^{\mu \nu \rho \lambda }\right)
.  \label{f1}
\end{equation}%
The collection indices from the three-form sector (capital, Latin
letters) are lowered with the (non-degenerate) metric $k_{AB}$
induced by the Lagrangian density $F_{\mu \nu \rho \lambda
}^{A}F_{A}^{\mu \nu \rho \lambda }$ in (\ref{f1}) (i.e. $F_{A}^{\mu
\nu \rho \lambda }=k_{AB}F^{B\mu \nu \rho \lambda }$) and are raised
with its inverse, of elements $k^{AB}$. The field strength of a
given three-form gauge field $V_{\mu \nu \rho }^{A}$ is defined in
the standard manner as $F_{\mu \nu \rho \lambda }^{A}=\partial
_{[\mu }V_{\nu \rho \lambda ]}^{A}$. Everywhere in this paper the
notation $[\mu \ldots \lambda ]$ signifies complete antisymmetry
with respect to the (Lorentz) indices between brackets, with the
conventions that the minimum number of terms is always used and the
result is never divided by the number of terms. The above action is
invariant under the gauge transformations
\begin{gather}
\delta _{\epsilon }A_{\mu }^{a}=\partial _{\mu }\epsilon ^{a},\quad \delta
_{\epsilon }\varphi _{a}=0,  \label{fx1} \\
\delta _{\epsilon }H_{\mu }^{a}=2\partial ^{\nu }\epsilon _{\mu \nu
}^{a},\quad \delta _{\epsilon }B_{a}^{\mu \nu }=-3\partial _{\rho
}\epsilon _{a}^{\mu \nu \rho },\quad \delta _{\epsilon }V_{\mu \nu
\rho }^{A}=\partial _{[\mu }^{\left. {}\right. }\epsilon _{\nu \rho
]}^{A},  \label{fx2}
\end{gather}%
where all the gauge parameters $\epsilon ^{a}$, $\epsilon _{\mu \nu }^{a}$,\
$\epsilon _{a}^{\mu \nu \rho }$\ and\ $\epsilon _{\mu \nu }^{A}$\ are
bosonic, with the last three sets completely antisymmetric. The gauge
algebra associated with (\ref{fx1}) and (\ref{fx2}) is Abelian.

We observe that if in (\ref{fx2}) we make the transformations
$\epsilon _{\mu \nu }^{a}\rightarrow \epsilon _{\mu \nu }^{a}\left(
\theta \right) =-3\partial ^{\rho }\theta _{\mu \nu \rho }^{a}$,\
$\epsilon _{a}^{\mu \nu \rho }\rightarrow \epsilon _{a}^{\mu \nu
\rho }\left( \theta \right) =4\partial _{\lambda }\theta _{a}^{\mu
\nu \rho \lambda }$,\ $\epsilon _{\mu \nu }^{A}\rightarrow \epsilon
_{\mu \nu }^{A}\left( \theta \right) =\partial _{[ \mu }^{\left.
{}\right. }\theta _{\nu ]}^{A}$,\ then the gauge variations from
(\ref{fx2}) identically vanish $\delta _{\epsilon \left( \theta
\right) }H_{\mu }^{a}\equiv 0$,\ $\delta _{\epsilon \left( \theta
\right) }B_{a}^{\mu \nu }\equiv 0$,\ $\delta _{\epsilon \left(
\theta \right)
}V_{\mu \nu \rho }^{A}\equiv 0$.\ Moreover, if we perform the changes $%
\theta _{\mu \nu \rho }^{a}\rightarrow \theta _{\mu \nu \rho }^{a}\left(
\phi \right) =4\partial ^{\lambda }\phi _{\mu \nu \rho \lambda }^{a}$,\ $%
\theta _{\mu }^{A}\rightarrow \theta _{\mu }^{A}\left( \phi \right)
=\partial _{\mu }\phi ^{A}$,\ with $\phi _{\mu \nu \rho \lambda }^{a}$\
completely antisymmetric functions and $\phi $\ an arbitrary scalar field,
then the transformed gauge parameters identically vanish $\epsilon _{\mu \nu
}^{a}\left( \theta \left( \phi \right) \right) \equiv 0$,\ $\epsilon _{\mu
\nu }^{A}\left( \theta \left( \phi \right) \right) \equiv 0$.\ Meanwhile,
there is no non-vanishing, local transformation of $\phi _{\mu \nu \rho
\lambda }^{a}$\ and $\phi ^{A}$\ that annihilates $\theta _{\mu \nu \rho
}^{a}\left( \phi \right) $\ and respectively $\theta _{\mu }^{A}\left( \phi
\right) $, and hence no further local reducibility identity. All these allow
us to conclude that the generating set of gauge transformations (\ref{fx1})
and (\ref{fx2}) is off-shell, second-order reducible.

The construction of the BRST symmetry for this free theory debuts with the
identification of the algebra on which the BRST differential $s$ acts. The
generators of the BRST algebra are of two kinds: fields/ghosts and
antifields. The ghost spectrum for the model under study comprises the
fermionic ghosts $\eta ^{\alpha _{1}}=\left( \eta ^{a},C_{\mu \nu }^{a},\eta
_{a}^{\mu \nu \rho },\bar{\eta}_{\mu \nu }^{A}\right) $ associated with the
gauge parameters $\left( \epsilon ^{a},\epsilon _{\mu \nu }^{a},\epsilon
_{a}^{\mu \nu \rho },\epsilon _{\mu \nu }^{A}\right) $ from (\ref{fx1}) and (%
\ref{fx2}), the bosonic ghosts for ghosts $\eta ^{\alpha _{2}}=\left( C_{\mu
\nu \rho }^{a},\eta _{a}^{\mu \nu \rho \lambda },\bar{\eta}_{\mu
}^{A}\right) $ due to the first-order reducibility parameters $\left( \theta
_{\mu \nu \rho }^{a},\theta _{a}^{\mu \nu \rho \lambda },\theta _{\mu
}^{A}\right) $, and also the fermionic ghost for ghosts for ghosts $\eta
^{\alpha _{3}}=\left( C_{\mu \nu \rho \lambda }^{a},\bar{\eta}^{A}\right) $
corresponding to the second-order reducibility parameters $\left( \phi _{\mu
\nu \rho \lambda }^{a},\phi ^{A}\right) $. The antifield spectrum is
organized into the antifields $\Phi _{\alpha _{0}}^{\ast }=( A_{a}^{\ast \mu
},H_{a}^{\ast \mu },\varphi ^{\ast a},B_{\mu \nu }^{\ast a},V_{A}^{\ast \mu
\nu \rho })$ of the original tensor fields and those corresponding to the
ghosts, denoted by $\eta _{\alpha _{1}}^{\ast }=\left( \eta _{a}^{\ast
},C_{a}^{\ast \mu \nu },\eta _{\mu \nu \rho }^{\ast a},\bar{\eta}_{A}^{\ast
\mu \nu }\right) $, $\eta _{\alpha _{2}}^{\ast }=\left( C_{a}^{\ast \mu \nu
\rho },\eta _{\mu \nu \rho \lambda }^{\ast a},\bar{\eta}_{A}^{\ast \mu
}\right) $, and respectively by $\eta _{\alpha _{3}}^{\ast }=\left(
C_{a}^{\ast \mu \nu \rho \lambda },\bar{\eta}_{A}^{\ast }\right) $. The
Grassmann parity of a given antifield is opposite to that of the associated
field/ghost.

The BRST\ symmetry of this free theory simply decomposes as the sum between
the Koszul-Tate differential $\delta $ and the exterior derivative along the
gauge orbits $\gamma $, $s=\delta +\gamma $, where the degree of $\delta $
is the antighost number ($\mathrm{antigh}\left( \delta \right) =-1$, $%
\mathrm{antigh}\left( \gamma \right) =0$), and that of $\gamma $ is the pure
ghost number ($\mathrm{pgh}\left( \gamma \right) =1$, $\mathrm{pgh}\left(
\delta \right) =0$). The grading of the BRST differential is named ghost
number ($\mathrm{gh}$) and is defined in the usual manner like the
difference between the pure ghost number and the antighost number, such that
$\mathrm{gh}\left( \delta \right) =\mathrm{gh}\left( \gamma \right) =\mathrm{%
gh}\left( s\right) =1$. According to the standard rules of the BRST method,
the corresponding degrees of the generators from the BRST complex are valued
like: $\mathrm{pgh}\left( \Phi ^{\alpha _{0}}\right) =0$, $\mathrm{pgh}%
\left( \eta ^{\alpha _{1}}\right) =1$, $\mathrm{pgh}\left( \eta ^{\alpha
_{2}}\right) =2$, $\mathrm{pgh}\left( \eta ^{\alpha _{3}}\right) =3$, $%
\mathrm{pgh}\left( \Phi _{\alpha _{0}}^{\ast }\right) =\mathrm{pgh}\left(
\eta _{\alpha _{1}}^{\ast }\right) =\mathrm{pgh}\left( \eta _{\alpha
_{2}}^{\ast }\right) =\mathrm{pgh}\left( \eta _{\alpha _{3}}^{\ast }\right)
=0$, $\mathrm{agh}\left( \Phi ^{\alpha _{0}}\right) =\mathrm{agh}\left( \eta
^{\alpha _{1}}\right) =\mathrm{agh}\left( \eta ^{\alpha _{2}}\right) =%
\mathrm{agh}\left( \eta ^{\alpha _{3}}\right) =0$, $\mathrm{agh}\left( \Phi
_{\alpha _{0}}^{\ast }\right) =1$, $\mathrm{agh}\left( \eta _{\alpha
_{1}}^{\ast }\right) =2$, $\mathrm{agh}\left( \eta _{\alpha _{2}}^{\ast
}\right) =3$, $\mathrm{agh}\left( \eta _{\alpha _{3}}^{\ast }\right) =4$.
The BRST differential is known to have a canonical action in a structure
named antibracket and denoted by the symbol $\left( ,\right) $ ($s\cdot
=\left( \cdot ,S\right) $), which is obtained by setting the fields/ghosts
respectively conjugated to the corresponding antifields. The generator of
the BRST symmetry is a bosonic functional of ghost number zero, which is
solution to the classical master equation $\left( S,S\right) =0$. The full
solution to the master equation for the free model under study reads as
\begin{eqnarray}
S& =&S_{0}+\int d^{4}x\left( A_{a}^{\ast \mu }\partial _{\mu }\eta
^{a}+2H_{a}^{\ast \mu }\partial ^{\nu }C_{\mu \nu }^{a}-3B_{\mu \nu
}^{\ast
a}\partial _{\rho }\eta _{a}^{\mu \nu \rho }\right. \notag \\
&& +V_{A}^{\ast \mu \nu \rho }\partial _{[\mu }^{\left. {}\right.
}\bar{\eta}_{\nu \rho ]}^{A}-3C_{a}^{\ast \mu \nu }\partial ^{\rho
}C_{\mu \nu \rho }^{a}+4\eta _{\mu \nu \rho }^{\ast a}\partial
_{\lambda
}\eta _{a}^{\mu \nu \rho \lambda } \notag \\
&& \left. +\bar{\eta}_{A}^{\ast \mu \nu }\partial _{[\mu }^{\left.
{}\right. }\bar{\eta}_{\nu ]}^{A}+4C_{\mu \nu \rho }^{\ast }\partial
_{\lambda }C^{\mu \nu \rho \lambda }+\bar{\eta}_{A}^{\ast \mu
}\partial _{\mu }\bar{\eta}^{A}\right) . \label{4}
\end{eqnarray}

Now, we consider the problem of constructing consistent interactions among
the fields $\Phi ^{\alpha _{0}}$ such that the couplings preserve the field
spectrum and the original number of gauge symmetries. The matter of
constructing consistent interactions is addressed by means of reformulating
this issue as a deformation problem of the solution to the master equation
corresponding to the free theory \cite{deflag}. Such a reformulation is
possible due to the fact that the solution to the master equation contains
all the information on the gauge structure of the theory. If an interacting
gauge theory can be consistently constructed, then the solution $S$ to the
master equation associated with the free theory can be deformed into a
solution $\bar{S}$
\begin{equation}
S\rightarrow \bar{S}=S+\lambda S_{1}+\lambda ^{2}S_{2}+\lambda
^{3}S_{3}+\cdots ,  \label{2.2}
\end{equation}%
of the master equation for the deformed theory
\begin{equation}
\left( \bar{S},\bar{S}\right) =0,  \label{2.3}
\end{equation}%
such that both the ghost and antifield spectra of the initial theory
are preserved. Equation (\ref{2.3}) splits, according to the various
orders in $\lambda $, into
\begin{align}
\left( S,S\right) & =0,  \label{2.4} \\
2\left( S_{1},S\right) & =0,  \label{2.5} \\
2\left( S_{2},S\right) +\left( S_{1},S_{1}\right) & =0,  \label{2.6} \\
& \vdots  \notag
\end{align}%
Equation (\ref{2.4}) is fulfilled by hypothesis. The next one
requires
that the first-order deformation of the solution to the master equation, $%
S_{1}$, is a co-cycle of the ``free" BRST
differential. However, only cohomologically non-trivial solutions to (\ref%
{2.5}) should be taken into account, as the BRST-exact ones (BRST
co-boundaries) correspond to trivial interactions. This means that
$S_{1}$ pertains to the ghost number zero cohomological space of
$s$, $H^{0}\left( s\right) $, which is generically non-empty due to
its isomorphism to the space of physical observables of the ``free"
theory. It has been shown (on behalf of the triviality of the
antibracket map in the cohomology of the BRST differential) that
there are no obstructions in finding solutions to the remaining
equations, namely (\ref{2.6}), etc.

The resolution of equations (\ref{2.5})--(\ref{2.6}), etc., implies
standard cohomological techniques related to the BRST differential
of the free model under consideration. In the sequel we give the
solutions to these equations without going into further details (to
be reported elsewhere). The (non-trivial) solution to equation
(\ref{2.5}) can be shown to expand like $S_{1}=\int d^{4}x\left(
\alpha _{0}+\alpha _{1}+\alpha _{2}+\alpha _{3}+\alpha _{4}\right)
$, where $\mathrm{antigh}\left( \alpha _{k}\right) =k $. The
component of antighost number four from the above decomposition
reads as
\begin{eqnarray}
\alpha _{4}&=&\left( P_{ab}\left( W\right) \right) ^{\mu \nu \rho
\lambda }\eta ^{a}C_{\mu \nu \rho \lambda }^{b}-\tfrac{1}{4}\left(
P_{ab}^{c}\left( M\right) \right) _{\mu \nu \rho \lambda }\eta
^{a}\eta ^{b}\eta _{c}^{\mu
\nu \rho \lambda } \notag \\
&&+Q_{aA}\left( f\right) \eta ^{a}\bar{\eta}^{A}+\tfrac{1}{4!}%
Q_{abcd}\left( f\right) \eta ^{a}\eta ^{b}\eta ^{c}\eta ^{d}+\tfrac{1}{2}%
Q^{ab}\left( f\right) \eta _{a\mu \nu \rho \lambda }\eta _{b}^{\mu
\nu \rho \lambda }, \label{bfa32}
\end{eqnarray}%
where we used the notations
\begin{eqnarray}
&&\left( P_{\Delta }\left( \chi \right) \right) ^{\mu \nu \rho \lambda }=%
\frac{\partial \chi _{\Delta }}{\partial \varphi _{a}}C_{a}^{\ast
\mu \nu \rho \lambda }+\frac{\partial ^{2}\chi _{\Delta }}{\partial
\varphi _{a}\partial \varphi _{b}}\left( H_{a}^{\ast [\mu
}C_{b}^{\ast \nu \rho \lambda ]}+C_{a}^{\ast [\mu \nu }C_{b}^{\ast
\rho \lambda
]}\right) \notag \\
&&+\frac{\partial ^{3}\chi _{\Delta }}{\partial \varphi _{a}\partial
\varphi _{b}\partial \varphi _{c}}H_{a}^{\ast [\mu }H_{b}^{\ast \nu
}C_{c}^{\ast \rho \lambda ]}+\frac{\partial ^{4}\chi _{\Delta
}}{\partial
\varphi _{a}\partial \varphi _{b}\partial \varphi _{c}\partial \varphi _{d}}%
H_{a}^{\ast \mu }H_{b}^{\ast \nu }H_{c}^{\ast \rho }H_{d}^{\ast
\lambda }, \label{pp}
\end{eqnarray}%
\begin{eqnarray}
&&Q_{\Lambda }\left( f\right) =f_{\Lambda }^{A}\bar{\eta}_{A}^{\ast
}-\left( P_{\Lambda }^{A}\left( f\right) \right) _{\mu
}\bar{\eta}_{A}^{\ast \mu
}-\left( P_{\Lambda }^{A}\left( f\right) \right) _{\mu \nu }\bar{\eta}%
_{A}^{\ast \mu \nu } \notag \\
&&+\left( P_{\Lambda }^{A}\left( f\right) \right) _{\mu \nu \rho
}V_{A}^{\ast \mu \nu \rho }-\tfrac{1}{4!}\left( P_{\Lambda
}^{A}\left( f\right) \right) _{\mu \nu \rho \lambda }F_{A}^{\mu \nu
\rho \lambda }, \label{q}
\end{eqnarray}%
and
\begin{align}
\left( P_{\Delta }\left( \chi \right) \right) ^{\mu \nu \rho }& =\frac{%
\partial \chi _{\Delta }}{\partial \varphi _{a}}C_{a}^{\ast \mu \nu \rho }+%
\frac{\partial ^{2}\chi _{\Delta }}{\partial \varphi _{a}\partial
\varphi _{b}}H_{a}^{\ast [\mu }C_{b}^{\ast \nu \rho
]}+\frac{\partial ^{3}\chi _{\Delta }}{\partial \varphi _{a}\partial
\varphi _{b}\partial \varphi _{c}}H_{a}^{\ast \mu }H_{b}^{\ast \nu
}H_{c}^{\ast \rho },  \label{p}
\\
\left( P_{\Delta }\left( \chi \right) \right) ^{\mu \nu }& =\frac{\partial
\chi _{\Delta }}{\partial \varphi _{a}}C_{a}^{\ast \mu \nu }+\frac{\partial
^{2}\chi _{\Delta }}{\partial \varphi _{a}\partial \varphi _{b}}H_{a}^{\ast
\mu }H_{b}^{\ast \nu },\quad \left( P_{\Delta }\left( \chi \right) \right)
^{\mu }=\frac{\partial \chi _{\Delta }}{\partial \varphi _{a}}H_{a}^{\ast
\mu }.  \label{p1}
\end{align}%
The functions $\left( P_{ab}\left( W\right) \right) ^{\mu \nu \rho \lambda }$
and $\left( P_{ab}^{c}\left( M\right) \right) _{\mu \nu \rho \lambda }$ are
obtained from (\ref{pp}) in which we replace $\chi _{\Delta }$ with $W_{ab}$
and respectively with $M_{ab}^{c}$, while the elements $Q_{aA}\left(
f\right) $, $Q_{abcd}\left( f\right) $, and $Q^{ab}\left( f\right) $ result
from the relations (\ref{q}) and (\ref{pp})--(\ref{p1}) where, instead of $%
f_{\Lambda }^{A}$, we put $f_{aB}^{A}$, $f_{abcd}^{A}$, and respectively $%
f^{Aab}$. (The objects $\left( P_{\Lambda }^{A}\left( f\right) \right) _{\mu
\nu \rho \lambda }$, $\left( P_{\Lambda }^{A}\left( f\right) \right) _{\mu
\nu \rho }$, $\left( P_{\Lambda }^{A}\left( f\right) \right) _{\mu \nu }$,
and $\left( P_{\Lambda }^{A}\left( f\right) \right) _{\mu }$ are expressed
by (\ref{pp}), (\ref{p}), and (\ref{p1}) in which $W_{\Delta }$ is
substituted by $f_{\Lambda }^{A}$). The quantities $W_{ab}$, $M_{ab}^{c}$, $%
f_{aB}^{A}$, $f_{abcd}^{A}$, and $f^{Aab}$ are arbitrary functions of the
undifferentiated scalar fields $\varphi _{a}$, with $M_{ab}^{c}$ and $%
f_{abcd}^{A}$ completely antisymmetric in their lower indices and $f^{Aab}$
symmetric in its BF indices. The piece of antighost number three from $S_{1}$
is given by
\begin{eqnarray}
&&\alpha _{3}=-\left( P_{ab}\left( W\right) \right) ^{\mu \nu \rho }\left(
\eta ^{a}C_{\mu \nu \rho }^{b}-4A^{a\lambda }C_{\mu \nu \rho \lambda
}^{b}\right) +2\left[ W_{ab}\eta _{\mu \nu \rho \lambda }^{\ast a}\right.
\notag \\
&&\left. +\left( P_{ab}\left( W\right) \right) _{[\mu \nu }B_{\rho \lambda
]}^{\ast a}+\left( P_{ab}\left( W\right) \right) _{[\mu }\eta _{\nu \rho
\lambda ]}^{\ast a}\right] C^{b\mu \nu \rho \lambda }-\left[ M_{ab}^{c}\eta
_{\mu \nu \rho \lambda }^{\ast a}\right.   \notag \\
&&\left. +\left( P_{ab}^{c}\left( M\right) \right) _{[\mu \nu }B_{\rho
\lambda ]}^{\ast a}+\left( P_{ab}^{c}\left( M\right) \right) _{[\mu }\eta
_{\nu \rho \lambda ]}^{\ast a}\right] \eta ^{b}\eta _{c}^{\mu \nu \rho
\lambda }-\left( Q^{ab}\left( f\right) \right) _{\mu }\eta _{a\nu \rho
\lambda }\eta _{b}^{\mu \nu \rho \lambda }  \notag \\
&&+\tfrac{1}{4}\left( P_{ab}^{c}\left( M\right) \right) _{\mu \nu \rho
}\left( \eta ^{a}\eta ^{b}\eta _{c}^{\mu \nu \rho }-8A_{\lambda }^{a}\eta
^{b}\eta _{c}^{\mu \nu \rho \lambda }\right) +\left( Q_{aA}\left( f\right)
\right) ^{\mu }\eta ^{a}\bar{\eta}_{\mu }^{A}  \notag \\
&&-\left[ \left( Q_{aA}\left( f\right) \right) ^{\mu }A_{\mu }^{a}+\left(
Q_{aA}\left( f\right) \right) ^{\mu \nu }B_{\mu \nu }^{\ast a}-\tfrac{1}{3}%
\left( Q_{aA}\left( f\right) \right) ^{\mu \nu \rho }\eta _{\mu \nu \rho
}^{\ast a}\right.   \notag \\
&&\left. -\tfrac{1}{12}\left( Q_{aA}\left( f\right) \right) ^{\mu \nu \rho
\lambda }\eta _{\mu \nu \rho \lambda }^{\ast a}\right] \bar{\eta}^{A}-\tfrac{%
1}{3!}\left[ \left( Q_{abcd}\left( f\right) \right) ^{\mu }A_{\mu
}^{a}+\left( Q_{abcd}\left( f\right) \right) ^{\mu \nu }B_{\mu \nu }^{\ast
a}\right.   \notag \\
&&\left. -\tfrac{1}{3}\left( Q_{abcd}\left( f\right) \right) ^{\mu \nu \rho
}\eta _{\mu \nu \rho }^{\ast a}-\tfrac{1}{12}\left( Q_{abcd}\left( f\right)
\right) ^{\mu \nu \rho \lambda }\eta _{\mu \nu \rho \lambda }^{\ast a}\right]
\eta ^{b}\eta ^{c}\eta ^{d},  \label{bfa33}
\end{eqnarray}%
where the functions appearing in the above and denoted by $\left(
Q^{ab}\left( f\right) \right) _{\mu }$, $\left( Q_{aA}\left(
f\right) \right) ^{\mu }$, $\left( Q_{aA}\left( f\right)
\right) ^{\mu \nu }$, $\left( Q_{aA}\left( f\right) \right) ^{\mu \nu \rho }$%
, $\left( Q_{aA}\left( f\right) \right) ^{\mu \nu \rho \lambda }$, $\left(
Q_{abcd}\left( f\right) \right) ^{\mu }$, $\left( Q_{abcd}\left( f\right)
\right) ^{\mu \nu }$, $\left( Q_{abcd}\left( f\right) \right) ^{\mu \nu \rho
}$, and $\left( Q_{abcd}\left( f\right) \right) ^{\mu \nu \rho \lambda }$
are withdrawn from the generic relations
\begin{eqnarray}
&&\left( Q_{\Lambda }\left( f\right) \right) ^{\mu }=-f_{\Lambda }^{A}\bar{%
\eta}_{A}^{\ast \mu }-2\left( P_{\Lambda }^{A}\left( f\right) \right) _{\nu }%
\bar{\eta}_{A}^{\ast \mu \nu }  \notag \\
&&+3\left( P_{\Lambda }^{A}\left( f\right) \right) _{\nu \rho }V_{A}^{\ast
\mu \nu \rho }-\tfrac{1}{3!}\left( P_{\Lambda }^{A}\left( f\right) \right)
_{\nu \rho \lambda }F_{A}^{\mu \nu \rho \lambda },  \label{q1}
\end{eqnarray}%
\begin{eqnarray}
\left( Q_{\Lambda }\left( f\right) \right) ^{\mu \nu } &=&2f_{\Lambda }^{A}%
\bar{\eta}_{A}^{\ast \mu \nu }-6\left( P_{\Lambda }^{A}\left( f\right)
\right) _{\rho }V_{A}^{\ast \mu \nu \rho }+\tfrac{1}{2}\left( P_{\Lambda
}^{A}\left( f\right) \right) _{\rho \lambda }F_{A}^{\mu \nu \rho \lambda },
\label{q2} \\
\left( Q_{\Lambda }\left( f\right) \right) ^{\mu \nu \rho } &=&-6f_{\Lambda
}^{A}V_{A}^{\ast \mu \nu \rho }+\left( P_{\Lambda }^{A}\left( f\right)
\right) _{\lambda }F_{A}^{\mu \nu \rho \lambda },  \label{q3} \\
\left( Q_{\Lambda }\left( f\right) \right) ^{\mu \nu \rho \lambda }
&=&-f_{\Lambda }^{A}F_{A}^{\mu \nu \rho \lambda }.  \label{q4}
\end{eqnarray}%
The remaining $P$-type coefficients from (\ref{bfa33}) result from relations
(\ref{p}) and (\ref{p1}). The last three constituents of $S_{1}$ are
expressed by the formulas
\begin{eqnarray}
&&\alpha _{2}=\left( P_{ab}\left( W\right) \right) ^{\mu \nu }\left( \eta
^{a}C_{\mu \nu }^{b}-3A^{a\rho }C_{\mu \nu \rho }^{b}\right) -2\left[ \left(
P_{ab}\left( W\right) \right) _{[\mu }B_{\nu \rho ]}^{\ast a}\right.   \notag
\\
&&\left. +W_{ab}\eta _{\mu \nu \rho }^{\ast a}\right] C^{b\mu \nu \rho }-%
\tfrac{1}{2}\left( P_{ab}^{c}\left( M\right) \right) _{\mu \nu }\left(
\tfrac{1}{2}\eta ^{a}B_{c}^{\mu \nu }+3A_{\rho }^{a}\eta _{c}^{\mu \nu \rho
}\right) \eta ^{b}  \notag \\
&&+\left[ \left( P_{ab}^{c}\left( M\right) \right) _{[\mu }B_{\nu \rho
]}^{\ast a}+M_{ab}^{c}\eta _{\mu \nu \rho }^{\ast a}\right] \eta ^{b}\eta
_{c}^{\mu \nu \rho }-\tfrac{1}{2}\left[ \left( P_{ab}^{c}\left( M\right)
\right) _{\mu }A_{c}^{\ast \mu }\right.   \notag \\
&&\left. -M_{ab}^{c}\eta _{c}^{\ast }\right] \eta ^{a}\eta ^{b}+\left\{ %
\left[ 3\left( P_{ab}^{c}\left( M\right) \right) _{\mu \nu }A_{\rho
}^{a}+12\left( P_{ab}^{c}\left( M\right) \right) _{\mu }B_{\nu \rho }^{\ast
a}\right. \right.   \notag \\
&&\left. \left. +4M_{ab}^{c}\eta _{\mu \nu \rho }^{\ast a}\right] A_{\lambda
}^{b}-6M_{ab}^{c}B_{\mu \nu }^{\ast a}B_{\rho \lambda }^{\ast b}\right\}
\eta _{c}^{\mu \nu \rho \lambda }+\tfrac{1}{2}\left( Q_{aA}\left( f\right)
\right) ^{\mu \nu }\left( \eta ^{a}\bar{\eta}_{\mu \nu }^{A}\right.   \notag
\\
&&\left. +A_{[\mu }^{a}\bar{\eta}_{\nu ]}^{A}\right) -\left( Q_{aA}\left(
f\right) \right) ^{\mu \nu \rho }B_{\mu \nu }^{\ast a}\bar{\eta}_{\rho }^{A}-%
\tfrac{1}{3}\left( Q_{aA}\left( f\right) \right) ^{\mu \nu \rho \lambda
}\eta _{\mu \nu \rho }^{\ast a}\bar{\eta}_{\lambda }^{A}  \notag \\
&&+\left[ \tfrac{1}{4}\left( Q_{abcd}\left( f\right) \right) ^{\mu \nu
}A_{\mu }^{a}A_{\nu }^{b}-\tfrac{1}{2}\left( Q_{abcd}\left( f\right) \right)
^{\mu \nu \rho }B_{\mu \nu }^{\ast a}A_{\rho }^{b}\right.   \notag \\
&&\left. -\tfrac{1}{4!}\left( Q_{abcd}\left( f\right) \right) ^{\mu
\nu \rho \lambda }\left( \eta _{[\mu \nu \rho }^{\ast a}A_{\lambda
]}^{b}-2B_{[\mu \nu }^{\ast a}B_{\rho \lambda ]}^{\ast b}\right)
\right]
\eta ^{c}\eta ^{d}  \notag \\
&&+\tfrac{1}{2}\left( Q^{ab}\left( f\right) \right) _{\mu \nu }\left(
-B_{a\rho \lambda }\eta _{b}^{\mu \nu \rho \lambda }+\tfrac{3}{4}\eta
_{a}^{\mu \alpha \beta }\eta _{b\;\alpha \beta }^{\nu }\right)   \notag \\
&&-\tfrac{1}{3}\left[ \left( Q^{ab}\left( f\right) \right) _{\mu \nu \rho
}A_{a\lambda }^{\ast }+\tfrac{1}{4}\left( Q^{ab}\left( f\right) \right)
_{\mu \nu \rho \lambda }\eta _{a}^{\ast }\right] \eta _{b}^{\mu \nu \rho
\lambda },  \label{bfa35}
\end{eqnarray}%
\begin{eqnarray}
&&\alpha _{1}=-\left( P_{ab}\left( W\right) \right) _{\mu }\left( \eta
^{a}H^{b\mu }-2A_{\nu }^{a}C^{b\mu \nu }\right) +W_{ab}\left( 2B_{\mu \nu
}^{\ast a}C^{b\mu \nu }\right.   \notag \\
&&\left. -\eta ^{a}\varphi ^{\ast b}\right) -\tfrac{1}{2}\left(
P_{ab}^{c}\left( M\right) \right) _{\mu }A_{\nu }^{a}\left( 2\eta
^{b}B_{c}^{\mu \nu }+3A_{\rho }^{b}\eta _{c}^{\mu \nu \rho }\right)   \notag
\\
&&-M_{ab}^{c}\left( B_{\mu \nu }^{\ast a}\eta ^{b}B_{c}^{\mu \nu }+A_{\mu
}^{a}\eta ^{b}A_{c}^{\ast \mu }+B_{[\mu \nu }^{\ast a}A_{\rho ]}^{b}\eta
_{c}^{\mu \nu \rho }\right)   \notag \\
&&-\tfrac{1}{3!}\left( Q_{aA}\left( f\right) \right) ^{\mu \nu \rho }\left(
A_{[\mu }^{a}\bar{\eta}_{\nu \rho ]}^{A}-\eta ^{a}V_{\mu \nu \rho
}^{A}\right) -\tfrac{1}{2}\left( Q_{aA}\left( f\right) \right) ^{\mu \nu
\rho \lambda }B_{\mu \nu }^{\ast a}\bar{\eta}_{\rho \lambda }^{A}  \notag \\
&&-\tfrac{1}{3!}\left[ \left( Q_{abcd}\left( f\right) \right) ^{\mu \nu \rho
}A_{\mu }^{a}A_{\nu }^{b}A_{\rho }^{c}+3\left( Q_{abcd}\left( f\right)
\right) ^{\mu \nu \rho \lambda }B_{\mu \nu }^{\ast a}A_{\rho }^{b}A_{\lambda
}^{c}\right] \eta ^{d}  \notag \\
&&+\tfrac{1}{4}\left( Q^{ab}\left( f\right) \right) _{\mu \nu \rho }\eta
_{a}^{\mu \nu \sigma }B_{b\sigma }^{\rho }-\tfrac{1}{12}\left( Q^{ab}\left(
f\right) \right) _{\mu \nu \rho \lambda }\eta _{a}^{\mu \nu \rho
}A_{b}^{\ast \lambda },  \label{bfa35x}
\end{eqnarray}%
and respectively
\begin{eqnarray}
&&\alpha _{0}=-W_{ab}A^{a\mu }H_{\mu }^{b}+\tfrac{1}{2}M_{ab}^{c}A_{\mu
}^{a}A_{\nu }^{b}B_{c}^{\mu \nu }-\tfrac{1}{4!}F_{B}^{\mu \nu \rho \lambda
}\left( f_{aA}^{B}A_{[\mu }^{a}V_{\nu \rho \lambda ]}^{A}+\right.   \notag \\
&&\left. +f_{abcd}^{B}A_{\mu }^{a}A_{\nu }^{b}A_{\rho }^{c}A_{\lambda }^{d}+%
\tfrac{1}{2}f^{Bab}B_{a\mu \nu }B_{b\rho \lambda }\right) .  \label{bf35y}
\end{eqnarray}%
In (\ref{bfa35}) and (\ref{bfa35x}) the functions of the type $P$
and $Q$ are yielded by formulas (\ref{p1}) and
(\ref{q2})--(\ref{q4}). Moreover, equation (\ref{2.5}) restricts the
functions $f_{aAB}$ to be antisymmetric in their three-form
collection indices,\ $f_{aAB}=-f_{aBA}$. This completes the general
form of the first-order deformation to the classical master
equation.

The second-order deformation (the solution to equation (\ref{2.6}))
can be shown to read as $S_{2}=\int d^{4}x\ \beta $, where
\begin{equation}
\beta =-\tfrac{1}{2\cdot 4!}\ H_{\mu \nu \rho \lambda }^{A}k_{AB}H^{B\mu \nu
\rho \lambda },  \label{bfa44}
\end{equation}%
with
\begin{eqnarray}
&&H_{\mu \nu \rho \lambda }^{B}=\left[ \left( P_{aA}^{B}\left( f\right)
\right) _{\mu \nu \rho \lambda }\eta ^{a}+\left( P_{aA}^{B}\left( f\right)
\right) _{[\mu \nu \rho }A_{\lambda ]}^{a}+2\left( P_{aA}^{B}\left( f\right)
\right) _{[\mu \nu }B_{\rho \lambda ]}^{\ast a}\right.   \notag \\
&&\left. +2\left( P_{aA}^{B}\left( f\right) \right) _{[\mu }\eta _{\nu \rho
\lambda ]}^{\ast a}+2f_{aA}^{B}\eta _{\mu \nu \rho \lambda }^{\ast a}\right]
\bar{\eta}^{A}-\left[ \left( P_{aA}^{B}\left( f\right) \right) _{[\mu \nu
\rho }\bar{\eta}_{\lambda ]}^{A}\right.   \notag \\
&&\left. -\left( P_{aA}^{B}\left( f\right) \right) _{[\mu \nu }\bar{\eta}%
_{\rho \lambda ]}^{A}-\left( P_{aA}^{B}\left( f\right) \right) _{[\mu
}V_{\nu \rho \lambda ]}^{A}\right] \eta ^{a}-\left[ \left( P_{aA}^{B}\left(
f\right) \right) _{[\mu \nu }A_{\rho }^{a}\bar{\eta}_{\lambda ]}^{A}\right.
\notag \\
&&\left. +2\left( P_{aA}^{B}\left( f\right) \right) _{[\mu }B_{\nu
\rho }^{\ast a}\bar{\eta}_{\lambda ]}^{A}+2f_{aA}^{B}\eta _{[\mu \nu
\rho }^{\ast a}\bar{\eta}_{\lambda ]}^{A}\right] -\left(
P_{aA}^{B}\left( f\right) \right) _{[\mu }A_{\nu
}^{a}\bar{\eta}_{\rho \lambda ]}^{A}  \notag
\\
&&-2f_{aA}^{B}B_{[\mu \nu }^{\ast a}\bar{\eta}_{\rho \lambda ]}^{A}+\tfrac{1%
}{3!}\left[ \tfrac{1}{4}\left( P_{abcd}^{B}\left( f\right) \right) _{\mu \nu
\rho \lambda }\eta ^{a}+\left( P_{abcd}^{B}\left( f\right) \right) _{[\mu
\nu \rho }A_{\lambda ]}^{a}\right.   \notag \\
&&\left. +2\left( P_{abcd}^{B}\left( f\right) \right) _{[\mu \nu }B_{\rho
\lambda ]}^{\ast a}+2\left( P_{abcd}^{B}\left( f\right) \right) _{[\mu }\eta
_{\nu \rho \lambda ]}^{\ast a}+2f_{abcd}^{B}\eta _{\mu \nu \rho \lambda
}^{\ast a}\right] \eta ^{b}\eta ^{c}\eta ^{d}  \notag \\
&&-\tfrac{1}{2}\left[ \left( P_{abcd}^{B}\left( f\right) \right) _{[\mu \nu
}A_{\rho }^{a}A_{\lambda ]}^{b}+2\left( P_{abcd}^{B}\left( f\right) \right)
_{[\mu }B_{\nu \rho }^{\ast a}A_{\lambda ]}^{b}\right.   \notag \\
&&\left. +2f_{abcd}^{B}\left( \eta _{[\mu \nu \rho }^{\ast
a}A_{\lambda ]}^{b}-2B_{[\mu \nu }^{\ast a}B_{\rho \lambda ]}^{\ast
b}\right) \right] \eta ^{c}\eta ^{d}-\left[ \left(
P_{abcd}^{B}\left( f\right) \right) _{[\mu }A_{\nu }^{a}A_{\rho
}^{b}A_{\lambda ]}^{c}\right.
\notag \\
&&\left. +2f_{abcd}^{B}B_{[\mu \nu }^{\ast a}A_{\rho }^{b}A_{\lambda ]}^{c}
\right] \eta ^{d}+\tfrac{1}{2}\left( P^{Bab}\left( f\right) \right) _{\mu
\nu \rho \lambda }\eta _{a\alpha \beta \gamma \delta }\eta _{b}^{\alpha
\beta \gamma \delta }  \notag \\
&&-\left( P^{Bab}\left( f\right) \right) ^{\alpha \beta \gamma }\eta
_{a\alpha \beta \gamma }\eta _{b\mu \nu \rho \lambda }+\tfrac{3}{8}\left(
P^{Bab}\left( f\right) \right) ^{\alpha \beta }\eta _{a\alpha \beta [\mu }\eta _{\nu \rho \lambda ]b}  \notag \\
&&+\left[ \left( P^{Bab}\left( f\right) \right) ^{\alpha \beta }B_{a\alpha
\beta }+2\left( P^{Bab}\left( f\right) \right) ^{\alpha }A_{a\alpha }^{\ast
}-2f^{Bab}\eta _{a}^{\ast }\right] \eta _{b\mu \nu \rho \lambda }  \notag \\
&&+\tfrac{1}{2}\left( P^{Bab}\left( f\right) \right) ^{\sigma
}B_{a\sigma [\mu }\eta _{\nu \rho \lambda
]b}-\tfrac{1}{2}f^{Bab}A_{a[\mu }^{\ast }\eta _{\nu \rho \lambda
]b}^{\left. {}\right. }+f_{aA}^{B}A_{[\mu
}^{a}V_{\nu \rho \lambda ]}^{A}  \notag \\
&&+f_{abcd}^{B}A_{\mu }^{a}A_{\nu }^{b}A_{\rho }^{c}A_{\lambda }^{d}+\tfrac{1%
}{3!}f^{Bab}B_{a[\mu \nu }B_{\rho \lambda ]b}.  \label{no4}
\end{eqnarray}%
In the meantime, equation (\ref{2.6}) requests that the various
functions depending on the undifferentiated scalar fields that
parameterize
the first-order deformation are subject to the equations%
\begin{gather}
W_{ea}\frac{\partial W_{bc}}{\partial \varphi _{e}}+W_{eb}\frac{\partial
W_{ca}}{\partial \varphi _{e}}+W_{ec}M_{ab}^{e}=0,\quad W_{e[a}\frac{%
\partial M_{bc]}^{d}}{\partial \varphi _{e}}+M_{e[a}^{d}M_{bc]}^{e}=0,
\label{i1} \\
-W_{e[a}\frac{\partial f_{b]B}^{A}}{\partial \varphi _{e}}%
-M_{ab}^{e}f_{eB}^{A}+f_{aE}^{A}f_{bB}^{E}-f_{bE}^{A}f_{aB}^{E}=0,\quad
f^{Aae}W_{eb}=0,  \label{i2} \\
W_{f[a}\frac{\partial f_{bcde]}^{A}}{\partial \varphi _{f}}%
+f_{f[abc}^{A}M_{de]}^{f}-f_{[abcd}^{E}f_{e]E}^{A}=0,\quad
f_{eC}^{(A}f_{\left. {}\right. }^{B)ae}=0,  \label{i5} \\
W_{ec}\frac{\partial f^{Aab}}{\partial \varphi _{e}}+f_{\left. {}\right.
}^{Ae(a}M_{ec}^{b)}-f_{cM}^{A}f^{Mab}=0,\quad f_{ebcd}^{(A}f_{\left.
{}\right. }^{B)ae}=0.  \label{i8}
\end{gather}%
Further, by direct computation we infer that $\left( S_{1},S_{2}\right) =0$,
so all the other deformations, of order three or higher, can be taken to
vanish, $S_{3}=S_{4}=\cdots =S_{k}=\cdots =0$. In conclusion, the full
deformed solution to the master equation for the model under study, which is
consistent to all orders in the coupling constants, can be written as $\bar{S%
}=S+\lambda \int d^{4}x\left( \alpha _{0}+\alpha _{1}+\alpha _{2}+\alpha
_{3}+\alpha _{4}\right) +\lambda ^{2}\int d^{4}x\ \beta $, where its
first-order components are listed in (\ref{bfa32}), (\ref{bfa33}), (\ref%
{bfa35})--(\ref{bf35y}) and $\beta $ is expressed by (\ref{bfa44}). From the
deformed solution to the master equation we can extract all the information
on the (gauge) structure of the resulting interacting model.

Thus, the piece of antighost number zero from the deformed solution $\bar{S}$
is precisely the Lagrangian action of the coupled model and has the
expression
\begin{equation}
\hat{S}[A_{\mu }^{a},H_{\mu }^{a},\varphi _{a},B_{a}^{\mu \nu },V_{\mu \nu
\rho }^{A}]=\int d^{4}x\left( H_{\mu }^{a}D^{\mu }\varphi _{a}+\tfrac{1}{2}%
B_{a}^{\mu \nu }\bar{F}_{\mu \nu }^{a}-\tfrac{1}{2\cdot 4!}\bar{F}_{\mu \nu
\rho \lambda }^{A}\bar{F}_{A}^{\mu \nu \rho \lambda }\right) ,  \label{def1}
\end{equation}%
where we employed the notations%
\begin{equation}
D_{\mu }\varphi _{a}=\partial _{\mu }\varphi _{a}+\lambda
W_{ab}A_{\mu }^{b},\quad \bar{F}_{\mu \nu }^{a}=\partial _{[\mu
}^{\left. {}\right. }A_{\nu ]}^{a}+\lambda M_{bc}^{a}A_{\mu
}^{b}A_{\nu }^{c}, \label{wza1}
\end{equation}%
\begin{eqnarray}
\bar{F}_{\mu \nu \rho \lambda }^{A} &=&F_{\mu \nu \rho \lambda }^{A}+\lambda
\left( f_{abcd}^{A}A_{\mu }^{a}A_{\nu }^{b}A_{\rho }^{c}A_{\lambda
}^{d}+f_{aB}^{A}A_{[\mu }^{a}V_{\nu \rho \lambda ]}^{B}\right.   \notag \\
&&\left. +\tfrac{1}{3!}f^{Aab}B_{a[\mu \nu }B_{\rho \lambda ]b}\right) .
\label{wza2}
\end{eqnarray}%
Under the general hypotheses mentioned at the beginning of this paper,
formula (\ref{def1}) gives the most general form of the action describing
the four-dimensional interactions between a collection of BF models and a
set of three-form gauge fields, whose free limit is (\ref{f1}). The action (%
\ref{def1}) is invariant under the deformed gauge transformations
\begin{equation}
\bar{\delta}_{\epsilon }\varphi _{a}=-\lambda W_{ab}\epsilon ^{b},\quad \bar{%
\delta}_{\epsilon }A_{\mu }^{a}=\left( D_{\mu }\right) _{\;\;b}^{a}\epsilon
^{b}+\tfrac{\lambda }{12}f_{A}^{ab}\bar{F}_{\mu \nu \rho \lambda
}^{A}\epsilon _{b}^{\nu \rho \lambda },  \label{def6}
\end{equation}%
\begin{eqnarray}
&&\bar{\delta}_{\epsilon }B_{a}^{\mu \nu }=-3\left( D_{\rho }\right)
_{a}^{\;\;b}\epsilon _{b}^{\mu \nu \rho }+\lambda \left( 2W_{ab}\epsilon
^{b\mu \nu }-M_{ab}^{c}B_{c}^{\mu \nu }\epsilon ^{b}\right)   \notag \\
&&+\tfrac{\lambda }{2}\bar{F}_{A}^{\mu \nu \rho \lambda }\left(
f_{abcd}^{A}A_{\rho }^{b}A_{\lambda }^{c}\epsilon ^{d}+f_{aB}^{A}\epsilon
_{\rho \lambda }^{B}\right) ,  \label{def8}
\end{eqnarray}%
\begin{eqnarray}
&&\bar{\delta}_{\epsilon }H_{\mu }^{a}=2\left( \tilde{D}^{\nu }\right)
_{\;\;b}^{a}\epsilon _{\mu \nu }^{b}-\tfrac{3\lambda }{2}\frac{\partial
M_{bc}^{d}}{\partial \varphi _{a}}A^{b\nu }A^{c\rho }\epsilon _{d\mu \nu
\rho }-\tfrac{\lambda }{12}\frac{\partial f^{Abc}}{\partial \varphi _{a}}%
B_{b\mu \alpha }\epsilon _{c\beta \gamma \delta }\bar{F}_{A}^{\alpha \beta
\gamma \delta }  \notag \\
&&-\tfrac{\lambda }{2}\bar{F}_{\mu \nu \rho \lambda }^{A}\left( \frac{%
\partial f_{bA}^{B}}{\partial \varphi _{a}}A^{b\nu }\epsilon _{B}^{\rho
\lambda }-\tfrac{1}{3}\frac{\partial f_{bA}^{B}}{\partial \varphi _{a}}%
\epsilon ^{b}V_{B}^{\mu \nu \rho }-\tfrac{1}{3}\frac{\partial f_{Abcde}}{%
\partial \varphi _{a}}A^{b\nu }A^{c\rho }A^{d\lambda }\epsilon ^{e}\right)
\notag \\
&&-\lambda \left( \frac{\partial W_{bc}}{\partial \varphi _{a}}H_{\mu }^{c}-%
\frac{\partial M_{bc}^{d}}{\partial \varphi _{a}}A^{c\nu }B_{d\mu \nu
}\right) \epsilon ^{b},  \label{def9}
\end{eqnarray}%
\begin{equation}
\bar{\delta}_{\epsilon }V_{\mu \nu \rho }^{A}=\left( D_{[\mu
}^{\left. {}\right. }\right) _{\;\;B}^{A}\epsilon _{\nu \rho
]}^{B}-\lambda f_{aB}^{A}V_{\mu \nu \rho }^{B}\epsilon ^{a}+\lambda
f_{abcd}^{A}A_{\mu }^{a}A_{\nu }^{b}A_{\rho }^{c}\epsilon
^{d}+\tfrac{\lambda }{2}f^{Aab}\sigma ^{\alpha \beta }B_{a\alpha
[\mu }\epsilon _{\nu \rho ]\beta b}, \label{def12}
\end{equation}%
where we used the notations
\begin{align}
\left( D_{\mu }\right) _{\;\;b}^{a}& =\delta _{b}^{a}\partial _{\mu
}-\lambda M_{bc}^{a}A_{\mu }^{c},\quad \left( D_{\mu }\right)
_{a}^{\;\;b}=\delta _{a}^{b}\partial _{\mu }+\lambda M_{ac}^{b}A_{\mu }^{c},
\label{deriv1a} \\
\left( \tilde{D}_{\mu }\right) _{\;\;b}^{a}& =\delta _{b}^{a}\partial _{\mu
}-\lambda \frac{\partial W_{bc}}{\partial \varphi _{a}}A_{\mu }^{c},\quad
\left( D_{\mu }\right) _{\;\;B}^{A}=\delta _{B}^{A}\partial _{\mu }+\lambda
f_{aB}^{A}A_{\mu }^{a}.  \label{deriv1}
\end{align}%
The gauge transformations (\ref{def6})--(\ref{def12}) remain second-order
reducible, but the reducibility relations only hold on-shell (where on-shell
means here on the stationary surface of the field equations for the action (%
\ref{def1})). These relations have an intricate, but not illuminating form,
and therefore we will skip them. The gauge algebra accompanying the deformed
gauge transformations (\ref{def6})--(\ref{def12}) is open, in contrast to
the original one, which is Abelian.

At this point, we have the entire information on the gauge structure of the
deformed theory. From (\ref{def1})--(\ref{wza2}) we observe that there
appear two main types of vertices. The first type, $\lambda \left( H_{\mu
}^{a}W_{ab}A^{b\mu }+\tfrac{1}{2}B_{b}^{\mu \nu }M_{ac}^{b}A_{\mu
}^{a}A_{\nu }^{c}\right) $, describes the self-interactions among the BF
fields in the absence of the three-forms and has been previously obtained in
the literature \cite{defBFijmpa,defBFjhep}. The second kind of vertices can
be put in the form%
\begin{eqnarray}
&&-\tfrac{\lambda }{4!}k_{AB}\partial ^{[\mu }V^{\nu \rho \lambda
]A}\left( f_{abcd}^{B}A_{\mu }^{a}A_{\nu }^{b}A_{\rho
}^{c}A_{\lambda
}^{d}+f_{aC}^{B}A_{[\mu }^{a}V_{\nu \rho \lambda ]}^{C}+\tfrac{1}{3!}%
f^{Bab}B_{a[\mu \nu }B_{\rho \lambda ]b}\right)   \notag \\
&&-\tfrac{\lambda ^{2}}{2\cdot 4!}k_{AB}\left( f_{abcd}^{A}A_{\mu
}^{a}A_{\nu }^{b}A_{\rho }^{c}A_{\lambda }^{d}+f_{aC}^{A}A_{[\mu }^{a}V_{\nu
\rho \lambda ]}^{C}+\tfrac{1}{3!}f^{Aab}B_{a[\mu \nu }B_{\rho \lambda
]b}\right)   \notag \\
&&\times \left( f_{emnp}^{B}A^{e\mu }A^{m\nu }A^{n\rho }A^{p\lambda
}+f_{eD}^{B}A^{e[\mu }V^{\nu \rho \lambda ]D}+\tfrac{1}{3!}%
f^{Bmn}B_{m}^{[\mu \nu }B_{n}^{\rho \lambda ]}\right) .  \label{secvert}
\end{eqnarray}%
We remark that (\ref{secvert}) contains some vertices involving only the BF
fields%
\begin{eqnarray}
&&-\tfrac{\lambda ^{2}}{2\cdot 4!}k_{AB}\left( f_{abcd}^{A}A_{\mu
}^{a}A_{\nu }^{b}A_{\rho }^{c}A_{\lambda }^{d}+\tfrac{1}{3!}f^{Aab}B_{a[\mu
\nu }B_{\rho \lambda ]b}\right)   \notag \\
&&\times \left( f_{emnp}^{B}A^{e\mu }A^{m\nu }A^{n\rho }A^{p\lambda }+\tfrac{%
1}{3!}f^{Bmn}B_{m}^{[\mu \nu }B_{n}^{\rho \lambda ]}\right) ,
\label{secvert1}
\end{eqnarray}%
whose existence is nevertheless induced by the presence of the three-form
gauge fields. Indeed, in the absence of these fields ($k_{AB}=0$) (\ref%
{secvert1}) vanishes. The remaining terms from (\ref{secvert}) produce
cross-couplings between the BF fields and the three-forms. From (\ref%
{secvert}) it is clear that the one-forms $H_{\mu }^{a}$ (from the BF
sector) cannot be coupled to the three-form gauge fields. The deformed gauge
transformations (\ref{def6})--(\ref{def12}) exhibit a rich structure, which
includes, among others, the generalized covariant derivatives (\ref{deriv1a}%
) and (\ref{deriv1}). It is interesting to notice that the presence of the
three-forms modifies the gauge transformations of $A_{\mu }^{a}$, $%
B_{a}^{\mu \nu }$, and $H_{\mu }^{a}$ by terms proportional with the
deformed field strength $\bar{F}_{\mu \nu \rho \lambda }^{A}$. Although the
one-forms $H_{\mu }^{a}$ do not couple to the three-form gauge fields, their
gauge transformations contain the gauge parameters $\epsilon _{B}^{\rho
\lambda }$, specific to the three-form sector. Similarly, the BF sector
contributes to the gauge transformations of the three-forms.

The previous results have been obtained in $D=4$ space-time dimensions. We
mention that the resulting cross-coupling terms originate in the pieces from
(\ref{bfa32}) proportional with $Q_{aA}\left( f\right) $, $Q_{abcd}\left(
f\right) $, and $Q^{ab}\left( f\right) $. These pieces are consistent
independently one from another (and also from the other terms present in (%
\ref{bfa32})) at the level of the first-order deformation. Let us consider
now the case $D>4$ (for $D<4$ the field strengths of the three-forms vanish,
such that no cross-couplings occur). In this case the gauge transformations (%
\ref{fx1})--(\ref{fx2}) from the BF sector are $\left( D-2\right)
$-order reducible. This implies the introduction of a larger
spectrum of ghosts and antifields for the BF sector than in $D=4$.
In addition,
the first-order deformation will accordingly stop at antighost number $D$, $%
S_{1}=\int d^{D}x(\alpha _{0}+\cdots +\alpha _{D})$. Standard cohomological
arguments can be used in order to establish that $\alpha _{D}$ will depend
only on the BRST generators from the BF sector. As a consequence, all the
components from the first-order deformation generated by $\alpha _{D}$ will
contribute only to pure BF couplings. On the other hand, basic cohomological
arguments ensure that the three-form BRST sector will occur non-trivially in
the first-order deformation only starting with terms of antighost number
four (just like in $D=4$) via the pieces from $\alpha _{4}$ proportional
with $Q_{aA}\left( f\right) $ and $Q_{abcd}\left( f\right) $ (see (\ref%
{bfa32}); the piece proportional with $Q^{ab}\left( f\right) $ is absent in $%
D>4$ due to the fact that the ghosts $\eta _{a\mu \nu \rho \lambda }$ are no
longer $\gamma $-invariant). Just like in $D=4$, the terms from $\alpha _{4}$
proportional to $Q_{aA}\left( f\right) $ and $Q_{abcd}\left( f\right) $ will
be consistent independently one from each other (and also from other pure BF
terms) and will yield the same results like in the case $D=4$. By contrast,
all the contributions coming from the term proportional with $Q^{ab}\left(
f\right) $ must be discarded from the first- and also from the second-order
deformations in $D>4$ (in particular, the term $\tfrac{1}{3!}%
f^{Aab}B_{a}^{[\mu \nu }B_{b}^{\rho \lambda ]}$ is absent from the deformed
action and accompanying gauge transformations). In conclusion, the
interacting action in $D>4$ will have a form similar to (\ref{def1}) up to
the fact that $\bar{F}^{A\mu \nu \rho \lambda }$ will lack the term $\tfrac{1%
}{3!}f^{Aab}B_{a}^{[\mu \nu }B_{b}^{\rho \lambda ]}$. In this situation the
deformed gauge transformations of $A_{\mu }^{a}$, $H_{\mu }^{a}$, and $%
V_{\mu \nu \rho }^{A}$ will no longer contain terms proportional with the
gauge parameters $\epsilon _{a}^{\mu \nu \rho }$. It is understood that the
deformed field strength $\bar{F}_{\mu \nu \rho \lambda }^{A}$ must be
replaced everywhere with its new expression, as explained in the above. The
previous discussion emphasizes that the case $D=4$ is a privileged situation
because it outputs the richest gauge structure for the cross-couplings
between the BF models and the three-forms.

Our procedure is consistent provided the equations (\ref{i1})--(\ref{i8})
are shown to possess solutions. In the sequel we give some classes of
solutions to these equations, without pretending to exhaust all their
possible solutions.

A first class of solutions is given by $M_{ab}^{c}=\frac{\partial W_{ab}}{%
\partial \varphi _{c}}$, $f_{aB}^{A}=k^{m}\lambda _{B}^{A}W_{am}$, $%
f_{abcd}^{A}=\mu ^{A}f_{e[ab}\frac{\partial W_{cd]}}{\partial \varphi _{e}}$%
, $f^{Aab}=0$, where $k^{m}$\ are some arbitrary constants and $\mu ^{A}$\
together with $\lambda _{B}^{A}$\ are some constants subject to the
conditions $\lambda _{B}^{A}\mu ^{B}=0$, while the non-degenerate matrix of
elements $W_{ab}$\ must satisfy the equations
\begin{equation}
W_{e[a}\frac{\partial W_{bc]}}{\partial \varphi _{e}}=0.  \label{z5}
\end{equation}%
We remark that all the non-vanishing solutions are in this case
parameterized by the antisymmetric functions $W_{ab}$.

We briefly review the basic notions on Poisson manifolds. If $P$\ denotes an
arbitrary Poisson manifold, then this is equipped with a Poisson bracket $%
\left\{ ,\right\} $\ that is bilinear, antisymmetric, subject to a
Leibnitz-like rule, and satisfies a Jacobi-type identity. If
$\left\{ X^{i}\right\} $\ are some local coordinates on $P$, then
there exists a two-tensor $\mathcal{P}^{ij}\equiv \left\{
X^{i},X^{j}\right\} $ (the Poisson tensor) that uniquely determines
the Poisson structure together with the Leibnitz rule. This
two-tensor is antisymmetric and transforms in a covariant manner
under coordinate transformations. Jacobi's identity for the Poisson
bracket $\left\{ ,\right\} $\ expressed in terms of the Poisson
tensor reads as
$\mathcal{P}_{,k}^{ij}\mathcal{P}^{kl}+\mathrm{cyclic}\left(
i,j,l\right) =0$, where $\mathcal{P}_{,k}^{ij}\equiv \partial \mathcal{P}%
^{ij}/\partial X^{k}$. In view of this discussion we can interpret the
functions $W_{ab}$ like the components of a two-tensor on a Poisson manifold
with the target space locally parameterized by the scalar fields $\varphi
_{e}$.

Another class of solutions to (\ref{i1})--(\ref{i8}) can be expressed as $%
W_{ab}=0$, $f_{aB}^{A}=0$, $f_{abcd}^{A}=0$, $f^{Aab}=\mu ^{ab}\xi ^{A}\hat{M%
}\left( \varphi \right) $, $M_{ab}^{c}=C_{\;\;ab}^{c}M\left( \varphi \right)
$, with $\hat{M}$\ and $M$\ arbitrary functions of the undifferentiated
scalar fields, $\xi ^{A}$\ some arbitrary constants, and $\mu ^{ab}$\ the
inverse of the Killing metric of a semi-simple Lie algebra with the
structure constants $C_{\;\;ab}^{c}$, where, in addition $C_{abc}=\bar{\mu}%
_{ad}C_{\;\;ab}^{d}$ (with $\bar{\mu}_{ad}\mu ^{de}=\delta _{a}^{e}$) must
be completely antisymmetric.

A third class of solutions can be written as $W_{ab}=0$, $f_{aB}^{A}=0$, $%
f^{Aab}=0$, $M_{ab}^{c}=\bar{C}_{\;\;ab}^{c}\hat{N}\left( \varphi \right) $,
$f_{abcd}^{A}=\bar{\xi}^{A}\bar{f}_{e[ab}\bar{C}_{\;\;cd]}^{e}N\left(
\varphi \right) $, where $\hat{N}$\ and $N$\ are some arbitrary functions of
the undifferentiated scalar fields, $\bar{\xi}^{A}$\ and $\bar{f}_{eab}$\
denote some arbitrary constants, and $\bar{C}_{\;\;ab}^{c}$\ are the
structure constants of a (in general not semi-simple) Lie algebra. Let us
particularize the last solutions to the case where $\bar{C}_{\;\;ab}^{c}=%
\bar{k}^{c}\bar{W}_{ab}$, $\hat{N}\left( \varphi \right) =N\left( \varphi
\right) =\frac{d\hat{w}\left( \bar{k}^{m}\varphi _{m}\right) }{d\left( \bar{k%
}^{n}\varphi _{n}\right) }$, with $\bar{k}^{c}$\ some arbitrary constants, $%
\hat{w}$\ an arbitrary, smooth function depending on $\bar{k}^{m}\varphi _{m}
$, and $\bar{W}_{ab}$\ some constants satisfying the relations $\bar{W}_{a[b}%
\bar{W}_{cd]}=0$. Obviously, the last relations ensure the Jacobi identity
for the structure constants $\bar{C}_{\;\;ab}^{c}$. Replacing back the
particular form of $\bar{C}_{\;\;ab}^{c}$, $\hat{N}$, and $N$ into the
initial solutions from the third class, we find $W_{ab}=0$, $%
f_{aB}^{A}=0=f^{Aab}$, $M_{ab}^{c}=\frac{\partial \hat{W}_{ab}}{\partial
\varphi _{c}}$, and $f_{abcd}^{A}=\bar{\xi}^{A}\bar{f}_{e[ab}\frac{\partial
\hat{W}_{cd]}}{\partial \varphi _{e}}$, where $\hat{W}_{ab}=\bar{W}_{ab}%
\frac{d\hat{w}\left( \bar{k}^{m}\varphi _{m}\right) }{d\left( \bar{k}%
^{n}\varphi _{n}\right) }$. It is easy to see, due to $\bar{W}_{a[b}\bar{W}%
_{cd]}=0$, that $\hat{W}_{ab}$\ satisfy the Jacobi identity for a Poisson
manifold, $\hat{W}_{e[a}\frac{\partial \hat{W}_{bc]}}{\partial \varphi _{e}}%
=0$. The above discussion emphasizes that we can generate solutions
correlated with a Poisson manifold even if $W_{ab}=0$. In this situation the
Poisson two-tensor results from a Lie algebra. It is interesting to remark
that the same equations, namely $\bar{W}_{a[b}\bar{W}_{cd]}=0$, ensure the
Jacobi identities for both the Lie algebra and the corresponding Poisson
manifold. These equations possess at least two types of solutions: $\bar{W}%
_{ab}=\varepsilon _{ijk}e_{a}^{i}e_{b}^{j}e_{c}^{k}\rho ^{c}$, with $i$,$j$,$%
k=\overline{1,3}$, and respectively $\bar{W}_{ab}=\varepsilon _{\bar{a}\bar{b%
}\bar{c}}l_{a}^{\bar{a}}l_{b}^{\bar{b}}l_{c}^{\bar{c}}\bar{\rho}^{c}$, with $%
\bar{a}$,$\bar{b}$,$\bar{c}=\overline{1,4}$, where $e_{a}^{i}$, $\rho ^{c}$,
$l_{a}^{\bar{a}}$, and $\bar{\rho}^{c}$\ are all constants and $\varepsilon
_{ijk}$\ together with $\varepsilon _{\bar{a}\bar{b}\bar{c}}$\ are
completely antisymmetric symbols, defined via the conventions $\varepsilon
_{123}=\varepsilon _{124}=\varepsilon _{134}=\varepsilon _{234}=+1$.

Finally, we consider the case where there exist some independent constants $%
\hat{k}^{a}$ such that $\hat{k}^{a}W_{ab}=0$, so $W_{ab}$ is degenerate. In
this situation a class of solutions to (\ref{i1})--(\ref{i8}) is expressed
by $M_{bc}^{a}=\frac{\partial W_{bc}}{\partial \varphi _{a}}$, $%
f_{aB}^{A}=k^{m}\lambda _{B}^{A}W_{am}$, $f_{abcd}^{A}=0$, and $f^{Aab}=\hat{%
k}^{a}\hat{k}^{b}\bar{M}\left( \hat{u}\right) $, where $\bar{M}$ is an
arbitrary function of $\hat{u}=\hat{k}^{a}\varphi _{a}$, $W_{ab}$ must
satisfy the Jacobi identity (\ref{z5}), $k^{m}$\ are some arbitrary
constants, and $\mu ^{A}$ are null vectors for $\lambda _{B}^{A}$. In this
case $W_{ab}$ depend on the undifferentiated scalar fields, but not
necessarily through $\hat{u}$. The solution $f^{Aab}$ is also degenerate
since it possesses the null vectors $\varepsilon _{bc}\left( \varphi
_{e}\right) \hat{k}^{c}$, with $\varepsilon _{bc}$ some antisymmetric,
arbitrary functions of the undifferentiated scalar fields.

In each of the four cases studied in the above the entire gauge structure of
the interacting model can be obtained by substituting the corresponding
solution into the formulas (\ref{def1})--(\ref{deriv1}). We remark that in
all these four cases we obtain interaction vertices among the BF fields
induced by the presence of the three-form gauge fields as well as vertices
that describe cross-couplings between BF fields and three-forms.

To conclude with, in this paper we have investigated the consistent
interactions that can be introduced between a collection of BF
theories and a set of three-form gauge fields. Starting with the
BRST differential for the free theory, we give the consistent
first-order deformation of the solution to the master equation, and
obtain that it is parameterized by several kinds of functions
depending only on the undifferentiated scalar fields. Next, we
determine the second-order deformation, whose existence imposes
certain restrictions with respect to these types of functions. Based
on these restrictions, we show that we can take all the remaining
higher-order deformations to vanish. As a consequence of our
procedure, we are led to an interacting gauge theory with deformed
gauge transformations, a non-Abelian gauge algebra that only closes
on-shell, and on-shell, second-order reducibility relations.
Finally, we investigate the equations that restrict the functions
parameterizing the deformed solution to the master equation, and
give some particular classes of solutions.

\section*{Acknowledgments}

The authors are partially supported by the European Commission FP6 program
MRTN-CT-2004-005104 and by the contract CEX 05-D11-49/07.10.2005 with the
Romanian Ministry of Education and Research (MEC).

\end{document}